\documentstyle[12pt,epsf,epsfig,psfig]{article}
\def\ga{\mathrel{\raise.3ex\hbox{$>$\kern-.75em\lower1ex\hbox{$\sim$}}}}
\def\la{\mathrel{\raise.3ex\hbox{$<$\kern-.75em\lower1ex\hbox{$\sim$}}}}
\def\gyr{{\rm \, G\kern-0.1em yr}}
\def\gev{{\rm \, Ge\kern-0.1em V}}
\def\tev{{\rm \, Te\kern-0.1em V}}
\def\beq{\begin{equation}}
\def\eeq{\end{equation}}

\def\mchi{m_{\tilde \chi}}

\def\m12{m_{1\!/2}}

\def\gf{\gamma_f}
\def\thm{\theta_\mu}
\def\tha{\theta_A}

\def\tb{\tan\beta}

\def\cp{C\!P}
\def\ohsq{\Omega_{\widetilde\chi}\, h^2}
\def\st{{\widetilde \tau}_{\scriptscriptstyle\rm R}}
\def\la{\stackrel{<}{{}_\sim}}
\def\ga{\stackrel{>}{{}_\sim}}
\newcommand{\be}{\begin{equation}}
\newcommand{\ee}{\end{equation}}

\newcommand{\s}{\mbox{$\sigma$}}

\newcommand{\fr}[2]{\frac{#1}{#2}}

\def\la{\stackrel{<}{{}_\sim}}
\def\ga{\stackrel{>}{{}_\sim}}
\begin{document} 
 
%%%%%%%%%%%%%%%% 
\begin{titlepage} 
\rightline{UMN--TH--1757/99}
\rightline{TPI--MINN--99/23}
\rightline{MADPH-99-1113}
\rightline{hep-ph/9904393}
\rightline{March 1999}  
\begin{center} 
 
\Large\bf MSSM Predictions for the Electric Dipole Moment of the 
$^{199}$Hg Atom 
\\ 
\medskip 
\vskip0.5in 
 
\normalsize   
 
{\bf Toby Falk}\footnote{falk@pheno.physics.wisc.edu}

{\it Department of Physics, University of Wisconsin, Madison, WI~53706, 
USA} 
 
{\bf Keith A. Olive}\footnote{olive@umn.edu}  and {\bf Maxim 
Pospelov}\footnote{pospelov@tpi1.hep.umn.edu}

\smallskip 
\medskip 
 
{ \it Theoretical Physics Institute, School of Physics and Astronomy, 
University of Minnesota, Minneapolis, MN 55455, USA} 

{\bf Radu Roiban}\footnote{roiban@insti.physics.sunysb.edu} 

{\it Institute for Theoretical Physics, State University of New York\\
Stony Brook, N. Y. 11794-3840, USA}  
 
\smallskip 
\end{center} 
\vskip1.0in 
 
\noindent{\large\bf Abstract} 
\smallskip\newline 
The Minimal Supersymmetric Standard Model can possess several CP-violating 
phases beyond the conventional Cabibo-Kobayashi-Maskawa phase. We calculate 
the contribution of these phases to T-violating nuclear forces. 
These forces induce a Schiff moment in the $^{199}$Hg nucleus,  
which is strongly limited by experiments aimed 
at the detection of the electric dipole moment of the mercury atom.  
The result for $d_{Hg}$ is found to be very 
sensitive to the CP-violating phases of the MSSM and the calculation
carries far fewer QCD uncertainties than the corresponding calculation of the
neutron EDM. In certain regions  of the MSSM parameter space, the limit
from the mercury EDM is  stronger than previous constraints based on
either the neutron  or electron EDMs. We present combined constraints
from the mercury  and electron EDMs to limit both CP-violating phases of
the MSSM.  We also present limits in mSUGRA models with unified gaugino
and scalar masses at the GUT scale. 
\end{titlepage} 
\setcounter{footnote}{0}
\section{Introduction} 
 
Despite the impressive success of the Standard Model, few are convinced that 
it is the final theory of particle interactions. For example, the 
supersymmetric modification of the Standard Model yields a very promising 
framework in which we are able to understand the stability of the 
electroweak scale. The Minimal Supersymmetric Standard Model (MSSM) provides 
a plethora of new phenomenological predictions which range from new charged 
and colored particles actively searched for in accelerators, to cold dark 
matter candidates, to new CP-violating phenomena such as the electric dipole 
moments of the neutron and electron which are generated if the additional 
CP-violating phases in the MSSM are non-zero. In this work, we study in 
detail the predictions of the MSSM for the electric dipole moment of the 
mercury atom and derive the constraints on the MSSM phases from the 
experimental limits on $d_{Hg}$. 
 
The null experimental results for the electric dipole moments (EDMs) of the 
electron, neutron, heavy atoms and diatomic molecules \cite 
{nEDM,mEDM,eEDM,molEDM} can in general place very strong constraints on the 
CP-violating sector of a new theory and probe energy scales which are 
inaccessible for direct observations at colliders \cite{KL}. In general, the 
relevant contribution to the dipole moments at scales of $\sim$1 GeV can be 
parameterized in terms of effective operators of different dimensions 
suppressed by corresponding powers of a high scale $M$ where these operators 
were generated:  
\begin{equation} 
{\cal L}_{eff}=\sum_{n\geq 4} \frac{c_{ni}}{M^{n-4}}{\cal O}^{(n)}_{i}, 
\end{equation} 
Here ${\cal O}_{i}^{(n)}$ are operators of dimension $n$, with its field 
content, Lorentz structures, etc., denoted by $i$. The fields relevant for the 
low-energy dynamics of interest are gluons, the three light quarks, the 
electron, and the electromagnetic field. This general form is independent of 
the particular construction of the new theory, and the details of a given 
model enter only through the values of the coefficients $c_{ni}/M^{n-4}$. 
 
In the MSSM, the number of operators which can generate an EDM is 
considerably smaller than in the generic case. In fact, all four-fermion 
operators are numerically insignificant. They can be generated in the MSSM 
only with additional factor of order $(m_{q}/M_{SUSY})^{2}$ modulo possible 
nontrivial flavor structure of the soft-breaking sector. Here we assume the 
minimal scenario with flavor-blind breaking of supersymmetry and therefore 
we can safely drop all  four-fermion CP-violating operators. Hence, 
the relevant part of the effective Lagrangian at the scale of 1 GeV contains 
the theta term, the three-gluon Weinberg operator, the EDMs of quarks and 
electron and the color EDMs (CEDMs) of quarks,  
\begin{eqnarray} 
{\cal L}_{eff} &=&\theta \frac{g_{s}^{2}}{32\pi ^{2}}G_{\mu \nu }^{a}
\mbox{$\tilde{G}$}_{\mu \nu }^{a}+w\frac{g_{s}^{3}}{6}f^{abc}G_{\mu \nu }^{a}
\mbox{$\tilde{G}$}_{\nu \alpha }^{b}G_{\alpha \mu }^{c}  \label{eq:eff} \\ 
&&+i\sum_{i=u,d,s}\frac{d_{i}}{2}\bar{q}_{i}F_{\mu \nu }\mbox{$\sigma$}_{\mu 
\nu }\mbox{$\gamma_{5}$}q_{i}+i\sum_{i=u,d,s}\frac{\tilde{d}_{i}}{2}\bar{q} 
_{i}g_{s}t^{a}G_{\mu \nu }^{a}\mbox{$\sigma$}_{\mu \nu }\mbox{$\gamma_{5}$} 
q_{i}+i\frac{\tilde{d}_{e}}{2}\bar{e}F_{\mu \nu }\mbox{$\sigma$}_{\mu \nu }
\mbox{$\gamma_{5}$}e. \nonumber
\end{eqnarray} 
 
We will assume here that the PQ mechanism of $\theta$-relaxation \cite{PQ} 
eliminates $\theta\sim O(1)$ and sets $\theta$ to $\theta_{eff}$ at the 
minimum of 
the axion potential. When both the CEDMs and Weinberg operator 
are absent, the value of $\theta_{eff}$ is exactly zero. However, nonzero $w$ 
and $\tilde d_i$ induce a linear term in the axion potential, and the 
effective value of $\theta$ is different from zero. This value leads to an 
additional contribution to the EDM of the neutron, usually ignored in the 
literature. 
 
The coefficients in front of the operators in Eq. (\ref{eq:eff}) can be 
calculated for any given model of CP-violation and then evolved down to 
the low-energy scale, using standard renormalization group techniques. 
In the MSSM, in particular, one can compute 
effective Lagrangian (\ref{eq:eff}) for any given point in the
supersymmetric parameter space.  
Then, to get the final predictions for EDMs, 
one has to take various matrix elements for these operators over hadronic, 
nuclear and atomic states \cite{KL,KKZ,KKY,KK}. In most cases this is a 
source of major uncertainty, especially when hadronic physics is involved. 
The exception is the case of a paramagnetic atom, in which the 
EDM is generated by 
the electron EDM $d_{e}$, and where the effects of 
nuclear CP-odd moments induced 
by the rest of the operators in (\ref{eq:eff}) can be safely neglected. 
The EDM of $^{205}$Tl is extremely sensitive to $d_{e}$ due to a 
very large relativistic enhancement factor $c\sim  -600$, which relates the 
EDM of the atom with $d_e$, $d_{Tl}=c d_e$. The experimental bound on 
the EDM of the thallium atom \cite{eEDM}, combined with good 
stability of atomic calculations (see \cite{KL} and 
references therein), leads to the 
following limit on the EDM of the electron:  
\begin{equation} 
d_{e}<4\cdot 10^{-27}e\cdot cm.  \label{de} 
\end{equation}

Therefore, the calculation of $d_{e}$ in the MSSM 
gives the most reliable limits
on CP-violating phases. It is clear, however, that the electron EDM limit 
alone cannot exclude the possibility of large CP-violating phases. This is 
because $d_{e}$, as any other coefficient in Eq. (\ref{eq:eff}), is in 
general a function of {\em several} CP-violating phases, and mutual 
cancellations are possible. This is what happens, for example, in the MSSM 
with the minimal number of parameters in the soft-breaking sector (see recent 
works \cite{Kane,FO1}). In the MSSM, it is well known that there are two 
independent CP-violating phases, $\theta_{\mu }$ and $\theta _{A}$, 
associated with the supersymmetric Higgs mass parameter $\mu $ and the soft 
supersymmetry breaking trilinear parameter $A$. The calculation of the 
relevant one loop diagrams determines $d_{e}$ as a function of these two 
phases. If the phases are small, $d_{e}$ is simply a linear combination of
$\theta _{A}$ and $\theta _{\mu }$. Therefore even a constraint as strong as
that  given in (\ref{de}) leaves a band on the $\theta _{A}$--$\theta _{\mu
}$  plane, along which a cancellation occurs and the phases are not
constrained.  In general, a second constraint could be expected to lift
this degeneracy and  place a strong constraint on both phases. It has
been common to use the limit on the neutron EDM as this second
constraint. Although there are large uncertainties in
the calculation of the neutron EDM, as we argue below,  when the limit on
the  neutron EDM is used, 
cancellations in the electron EDM occur in many of the same regions as
cancellations in the neutron EDM.  Therefore, one is led to the conclusion
that large phases are still possible. 
 
In what follows, we critically reexamine the reliability of 
the calculation of the EDM of the neutron in the MSSM. We demonstrate that 
this calculation is subject to very large hadronic uncertainties, which 
makes the extraction of the limits on CP-violating phases in MSSM 
tenuous. Instead, we propose that useful limits may be obtained from the 
limits on the EDM of the mercury atom. This arises from the T-odd 
nucleon-nucleon interaction in the MSSM, induced mainly due to the CEDMs of 
light quarks. This interaction gives rise to an EDM of the mercury atom by 
inducing the Schiff moment of mercury nucleus. We demonstrate that the 
degree of QCD uncertainties related to this calculation is in fact 
smaller than in the case of $d_n$ and that it is possible to calculate the 
T-odd nucleon-nucleon interaction as a function of the different MSSM 
phases. As an example, we proceed with the calculation of the EDM of the 
mercury atom in one specific point of the supersymmetric parameter space 
where all squark, gaugino masses, $|\mu|$ and $|A|$ parameters are set 
equal. This ``pilot'' calculation demonstrates the  sensitivity of
$d_{Hg}$ to the CP-violating phases of MSSM. We find in this case that
$d_{Hg}$ provides somewhat better limits on CP-violating phases than $d_e$.  
We proceed further and
combine mercury EDM and  electron EDM constraints to exclude most of the
parameter space on
$\theta_A$-- 
$\theta_\mu$ plane in this toy example. Finally, we consider more realistic 
constraints over the supersymmetric plane when supersymmetry 
breaking scalar and gaugino masses are unified at the GUT scale. In this 
case, we find that the limits on CP-violating phases obtained from
$d_{Hg}$ is no longer more restrictive than $d_e$, 
as the RG evolution of soft-breaking parameters makes squarks and gluino
heavier than sleptons, charginos and neutralinos. 
The combined limits are still very powerful as the cancellation of
different supersymmetric contributions typically occur in different regions
of parameter space.
 
\section{The Neutron EDM in the MSSM} 
 
Limits on the neutron EDM are commonly used to set constraints on new CP 
violating interactions. In particular, the upper limit to $d_{n}$ is often 
used to limit the size of the CP-violating phases in the MSSM \cite{ko}. 
The current  experimental limit on the EDM of the neutron is  
\begin{equation} 
d_{n}<1.1\cdot 10^{-25}e\cdot cm, 
\end{equation} 
Indeed, the EDM of the neutron receives contribution from all operators 
listed above in Eq. (\ref{eq:eff}) except $d_{e}$. However, there is a 
complication in using the neutron EDM as compared to the electron EDM, due to 
QCD uncertainties which make the extraction of the limits on CP-violating 
phases in the fundamental Lagrangian problematic. We demonstrate two aspects 
of this problem below. 
 
The most straightforward contribution to the EDM of the neutron is due to 
the quark EDM operators. It is usually estimated using nonrelativistic 
SU(6) quark model. The result,  
\begin{equation} 
d_{n}=\frac{4}{3}d_{d}-\frac{1}{3}d_{u},  \label{eq:naive} 
\end{equation} 
can be compared, in fact, with the model calculations \cite{MP} and lattice 
simulations of light quark tensor charges in the nucleon \cite{ADHK}. The 
matrix elements for the tensor charges of the nucleon are defined by  
\begin{equation} 
\langle N|{\bar{\psi}_{q}}\sigma _{\mu \nu }\psi _{q}|N\rangle =\delta q 
\overline{N}\sigma _{\mu \nu }N,
\end{equation} 
whereas the axial charges are defined by  
\begin{equation} 
\langle N|{\bar{\psi}_{q}}\gamma _{\mu }\gamma _{5}\psi _{q}|N\rangle 
=\Delta q\overline{N}~\gamma _{\mu }\gamma _{5}N 
\end{equation} 
In the Na\"{\i }ve quark model, both Lorentz structures correspond to the 
spin of a nonrelativistic quark. In this case $\delta u=\Delta u=-1/3$,   
$ \delta d=\Delta d=4/3$, $\delta s=\Delta s=0$, yielding  eq. 
(\ref{eq:naive}). Note that isospin symmetry gives us
$(\Delta u)_{n}=(\Delta 
d)_{p}$, etc. However, as argued in \cite{EF}, since it appears 
that the contribution to the nucleon spin from the strange quark ($\Delta 
_{s}$) is non-vanishing \cite{EMC},
 the na\"{\i }ve quark model may not be sufficient to describe the 
quark EDM contribution to the neutron EDM. While it is not the axial charges 
which need to be considered for the calculation of the neutron EDM, but
rather the tensor charges, the 
departure of the axial charge values from their NQM values indicates that 
more realistic (non-NQM) values of the tensor charges ($\delta q$) must be
used. According  to calculations based on Lattice QCD \cite{ADHK}, the
tensorial charges  of up and down quarks in the proton
%the coefficients in (\ref {eq:naive}) 
should be read as $\delta u\sim 0.8$ and $\delta d\sim -0.23$. 
This means that the naive nonrelativistic formula predicts the EDM of the 
neutron due to the quark EDMs to be 1.5-1.7 times larger than the lattice 
result. Slightly different values of $\delta u\sim 1.1$ and $\delta d\sim 
-0.4$ can be derived from the SU(3) chiral quark soliton model \cite{MP}. 
The tensor charge 
of the strange quark is found to be consistent with zero in both methods  
\cite{MP,ADHK}. This is due to the fact that the $\bar{s}\mbox{$\sigma$} 
_{\mu \nu }s$ operator is odd under charge conjugation which must result in 
the Zweig-type suppression of this matrix element over the neutron state. 
Even with the usual $m_{s}/m_{d}$ enhancement of this operator, it is 
unlikely to be important. This does not exclude other possible CP-violating 
operators involving the $s$-quark, CEDM or generic four-fermion operators, as 
their contributions to the EDM of the neutron can be significant \cite 
{KKZ,HPs}. Departures from the predictions of the non-relativistic quark
model were recently considered in \cite{bartl}.
 
Unfortunately, the quantitative evaluation of the remaining contributions to 
the neutron EDM is complicated due to of our lack of knowledge about strong 
interaction dynamics at 1 GeV and below. Typically, one resorts to Naive 
Dimensional Analysis (NDA) \cite{NDA}, formulated within the 
constituent quark framework. This method is, however, only 
an order of magnitude estimate to be used when other methods of calculation 
fail to produce an answer. When the problem of estimating $d_N$ due to 
$\tilde d_{u,d}$ is considered, there are several possible answers in the 
literature:
\begin{eqnarray}
d_N\simeq e0.7(\tilde d_u + \tilde d_d)& 
{\rm Ref.} \cite{KK}\label{cle}\\
d_N\sim\fr{eg_s}{4\pi}(O(1)\tilde d_u + O(1)\tilde d_d) &{\rm NDA, Ref. }
\cite{Duff}\label{duffe}
\end{eqnarray}
We have chosen a normalization where $g_s$ is included 
in the definition of the operator in (\ref{eq:eff}) and 
correspondingly include an
additional $g_s$ in the estimate (\ref{duffe}).
The first result is based on a combination of chiral perturbation 
theory and QCD sum rules. The latter estimate
is derived with the use of NDA \footnote{We note that the estimate in
\cite{Barbieri} is suppressed by an additional factor of $g_s/4\pi$.}.  For
a realistic choice of the strong coupling constant at the scale of $1$ GeV,
$g_s\sim
\sqrt{0.5\cdot 4\pi}=2.5$, the overall numerical coefficient in  eq.
(\ref{duffe})  is about 3.6 times  smaller than in (\ref{cle}). Estimates 
based on NDA imply that for natural relations among
coefficients $d_{i}/e\sim \tilde{d}_{i}$, the effects of color EDMs on the 
electric dipole moment of the neutron are negligible and the result can
indeed be approximated by the linear combination of EDMS of quarks. 

In fact, it is possible to show that the CEDMs can lead to a substantially 
larger contribution to the neutron EDM than some of the
predictions based on 
NDA. The easiest 
way to see that CEDMs can be  numerically important is to calculate the 
effective $\theta$-term induced by CEDMs in the presence of the PQ
symmetry and then use the result for $d_N(\theta)$. 
This value, $\theta_{eff}(\tilde d_i)$ can be calculated 
within the current algebra approach, in a manner similar to the calculation of the
vacuum topological susceptibility \cite{SVZ,BU,Posp}. The dynamically 
induced theta term can be expressed in the following compact form:  
\begin{equation} 
\theta_{eff}=-\frac{m^2}{2}\left(\frac{\tilde d_u}{m_u}+ \frac{\tilde d_d}
{m_d}+\frac{\tilde d_s}{m_s}\right).  \label{eq:theta} 
\end{equation} 
Here, $m^2$ is the ratio of the quark-gluon condensate to the quark 
condensate. It is known to good accuracy from QCD sum rules 
\cite{SR} that,  
\begin{equation} 
m^2=\frac{\langle 0|g \bar q (G\sigma)q|0\rangle}{\langle 0| \bar q 
q|0\rangle}\simeq 0.8\mbox{GeV}^2.  \label{eq:m0} 
\end{equation} 
 
The accuracy of the estimate (\ref{eq:theta}) is of order 
$m^2_{\pi,K}/m^2_{\eta^{\prime}}$, which is acceptable for our purposes. 
If no interference with other terms is expected, then the expression (\ref 
{eq:theta}) must be less than the current limit on $\theta$, extracted from 
the same neutron EDM data. Using the fact that in the simplest variant of the
MSSM, $\tilde d_d/m_d=\tilde d_s/m_s$, and assuming for a moment that this
is  the only contribution to the EDM of the neutron, one can obtain the 
following, quite stringent, level of sensitivity for the CEDM:  
\begin{equation} 
\tilde d_d<10^{-25} cm.  \label{eq:thelim} 
\end{equation} 
This fact alone suggests that CEDMs may contribute significantly to 
the EDM of the neutron, typically at the level of the prediction (9) and
an order of magnitude above NDA 
predictions.  Remarkably, the main uncertainty in the limit 
(\ref{eq:thelim}) comes not 
from  the calculation of $\theta(CEDM)$, but rather from the 
principal difficulties 
in calculating $d_N(\theta)$. In the standard approach \cite{CDVW}, the 
chiral loop diagram is used to estimate $d_N(\theta)$. This loop is 
logarithmically divergent in the exact chiral limit and therefore is 
distinguished from the rest of the contributions. For realistic values of the  
parameters, however, this logarithm is not large and other contributions can 
be equally important. This makes the whole calculation problematic even
in  predicting the sign of the $\theta$ term contribution to $d_N$.

Besides $d_N(\theta(CEDM))$, one should also consider direct CEDM-induced 
contributions to the EDM of the neutron which can be computed within the same 
chiral loop approach \cite{KK}. Combining different contributions, we can 
symbolically write the result for the EDM of the neutron in the following 
form:  
\begin{equation} 
d_N\simeq 0.8 d_d- 0.23 d_u + e\left[\tilde d_u \left(c_1\ln\frac{\Lambda 
}{m_\pi}+c_2\right)+\tilde d_d \left(c_3\ln\frac{\Lambda}{m_\pi} 
+c_4\right)+ \tilde d_s\left(c_5\ln\frac{\Lambda}{m_K}+c_6\right)\right]. 
\label{eq:symbolic} 
\end{equation} 
The coefficients $c_1$, $c_3$ and $c_5$ were estimated in Ref. \cite{KK} to 
be $c_1\ln(m_\rho/m_\pi)=c_3\ln(m_\rho/m_\pi)\simeq 0.7$ and 
$c_5\simeq 0.1$. The cutoff parameter $\Lambda$ 
corresponds to scales where chiral perturbation theory breaks down, that is,
of order $m_\rho$. In the exact chiral limit, $m_\pi,~m_K\rightarrow 0$, and 
the logarithmic terms dominate. In practice, however, the logarithmic terms 
are numerically not distinguished from the coefficients $c_2$, $c_4$ and 
$c_6$, which are {\em a priori} comparable with $c_1$, $c_3$ and $c_5$ and are 
not calculable in this approach. It is clear then that these terms can 
change both the value and the signs of different contributions to $d_N$. 
Therefore, although in principle very important as an order of magnitude 
estimate, Eq. (\ref 
{eq:symbolic}) fails to provide $d_N$ as a known function of individual 
$\tilde d_i$-contributions and, ultimately, of different CP-violating phases.

 As emphasized in Ref. \cite{W}, 
the NDA estimate of $d_n(\theta)$ essentially reproduces the  calculation
of Ref. \cite{CDVW}. The source of the disagreement in the case of 
$d_n(CEDMs)$ can be traced to the problem of estimating the CP-odd 
$\pi^+pn$--vertex, proportional to the 
matrix element $\langle p|\bar u g_s(G\s)d|n\rangle$. 
In Ref. \cite{KK} this
matrix element was estimated to be -1.5 GeV$^2$ and is essentially
proportional  to the quark-gluon condensate parameter $m^2\sim 0.8{\rm
GeV}^2$  (\ref{eq:m0}). 
On the other hand, it can be shown 
that NDA suggests for this matrix element a value of order $4\pi f_\pi^2\sim 
{\rm GeV}^2/(4\pi)$, i.e. one order of magnitude smaller. 
This difference is related to the fact that the NDA assumes 
nonrelativistic quarks whose chromomagnetic interactions are suppressed,
whereas QCD sum rules use more realistic descriptions of hadronic 
properties in terms of vacuum quark--gluonic condensates. 
 
To summarize this discussion, the extraction of reliable limits on the 
CP-violating phases in the MSSM from the EDM of the neutron is difficult and 
uncertain. Even the best estimates of $d_n$ 
based on the ``chiral logarithm'' 
approach \cite{KK}, bear a large degree of uncertainty and 
cannot produce a precise prediction for $d_n$ as a  function of the
CP-violating phases. Useful limits are still available from the electron
EDM; however,  the magnitude of the phases is not terribly constrained on
this basis alone,  due to cancellations in the various MSSM contributions
to $d_e$.  Fortunately, the EDM of the neutron is not the only source of
information  about CP-violation in the strongly-interacting sector.
Limits on T-violating nuclear forces are provided by experiments
aimed at the detection  of the EDM of paramagnetic atoms, among which the
EDM of  $^{199}$Hg atom is the most constraining. In what follows, we
will discuss the  constraints these limits provide, both alone and in
conjunction with the  electron EDM limits. 
 
\section{CP-violating nucleon-nucleon interaction in MSSM} 
 
The limits on T-odd nuclear forces extracted from the atomic experiments are 
in general very important for particle physics \cite{KL}. In the case of 
diamagnetic atoms, the most impressive limit is obtained for the EDM of 
$^{199}$Hg \cite{mEDM}:  
\begin{equation} 
d_{Hg}<9\cdot 10^{-28} e \cdot cm.  \label{eq:Hglim} 
\end{equation} 
The electric screening of the electric dipole moments of the atom's 
constituents is violated by the finite size of the nucleus and can be 
conveniently expressed by the Schiff moment $S$, which 
parametrizes the effective interaction between the electron and 
nucleus of spin ${\bf I}$, $V_{eff}=-eS({\bf I \nabla})\delta({\bf r})$ 
\cite{KL}. 
Atomic calculations derive the atomic EDM as a function of $S$ and 
translate the experimental result (\ref{eq:Hglim})  into the limit on the 
Schiff moment of the nucleus:  
\begin{eqnarray} 
d_{Hg}=S\cdot 3.2\cdot 10^{-18}\mbox{fm}^{-2}  \nonumber \\ 
S<2.8\cdot 10^{-10}e\mbox{fm}^3.  \label{eq:Sexp} 
\end{eqnarray} 
The Schiff moment of the nucleus can be induced either due to the Schiff moment of 
the valence nucleons or due to the breaking of time invariance in the 
nucleon-nucleon interaction, the latter being enhanced by the collective 
effects in the nucleus. The calculation of the Schiff moment of the nucleus, 
originating from various $\bar NN \bar N^{\prime}i\gamma_5N^{\prime}$ 
interactions was done in the single particle approximation with square-well 
and Woods-Saxon potentials \cite{KSF}. The results show that the Schiff 
moment of mercury is primarily sensitive to the $\bar pp \bar n i\gamma_5 n$ 
interaction. If we parameterize the coefficient in front of this interaction 
as $\xi G_F/\sqrt{2}$, the nuclear calculation \cite{KSF} provides us with the 
following value for $S$:  
\begin{equation} 
S=-1.8\cdot 10^{-7}\xi e \cdot\mbox{fm}^3.  \label{eq:Sth} 
\end{equation} 
Combined with Eq. (\ref{eq:Sexp}), it gives the following constraint on 
$\xi$:  
\begin{equation} 
\xi<1.9\cdot 10^{-3}. 
\end{equation} 
Questions concerning the
calculation of the strength of $\bar pp \bar n i\gamma_5 n$ interaction 
induced by different operators was considered in (\ref{eq:eff}),
\cite{KL,KKY,KK}.  The effective theta term, the Weinberg three-gluon 
operator and the CEDMs of quarks can generate this interaction. 
Numerically, the contributions provided by the CEDMs of up and down quarks
are  the most important and we concentrate our analysis on them, trying to 
incorporate the effect of $\tilde d_s$ as well. 
\begin{figure}[thb]
 \begin{center}
\epsfig{file=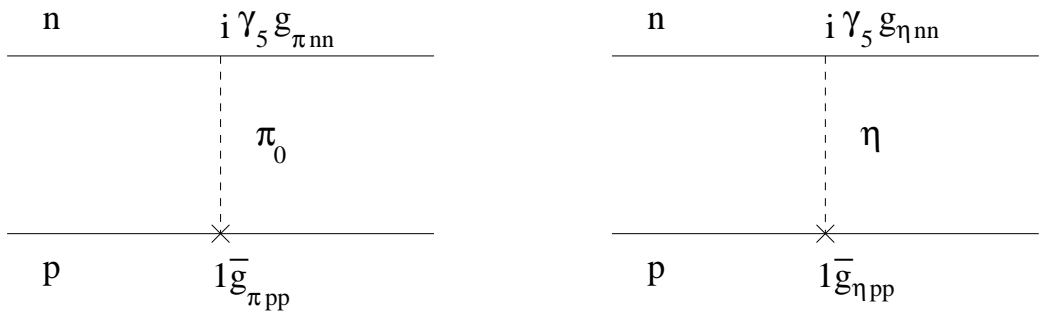,height=3in} 
\vspace{-1.5in}
\caption{Pseudoscalar meson exchange diagrams, inducing 
$\bar pp \bar n i \gamma_5 n $ 
interaction. }   \ \end{center}
\vspace{-0.3in}\end{figure}
 
Following \cite{KKY,KK}, we approximate the T-violating nucleon-nucleon 
interaction by pseudoscalar exchange, as shown in Fig 1. In the limit of 
exact chiral symmetry this exchange has the power-like singularity $ 
m_\pi^{-2}$, to be compared with the logarithmic singularity in the case 
of the EDM 
of the neutron. The CP violation resides in proton-meson vertex which can be 
calculated with QCD sum rules and current algebra 
techniques. The CP-conserving meson-neutron vertex is sufficiently well 
known from SU(3)-relations in baryon octet decay amplitudes and from the 
axial charges of nucleons. If only $\tilde d_u$ and $\tilde d_d$ are 
present, the pion exchange dominates $\eta$ exchange by a factor 
$m_\eta^2/m_\pi^2\simeq 16$. In the MSSM, though, the strange quark CEDM is 
enhanced relative to that of the down quark by a factor $m_s/m_d$ and 
$\eta$ meson exchange is not {\em a priori} negligible. In the chiral 
approach, CP-violating vertices of interest can be reduced to the following 
set of matrix elements:  
\begin{eqnarray} 
\bar g_{\pi pp} =\frac{\tilde d_u + \tilde d_d}{4f_\pi} \left(
\langle p|\bar u 
g_s(G\mbox{$\sigma$})u - \bar d g_s(G\mbox{$\sigma$})d |p\rangle \right) + 
\nonumber\\
\frac{\tilde 
d_u - \tilde d_d}{4f_\pi}\left(\langle p|\bar u g_s(G\mbox{$\sigma$})u 
+\bar d g_s(G \mbox{$\sigma$})d |p\rangle -
m^2\langle p|\bar u u +\bar d d |p\rangle\right) \nonumber \\ 
\bar g_{\eta pp} =-\frac{\tilde d_s}{\sqrt{3}f_\pi} 
\left (\langle p|\bar s g_s(G\mbox{$\sigma$})s|p\rangle - 
m^2\langle p|\bar s s|p\rangle\right)
\label{eq:matrel} 
\end{eqnarray} 
Here $m^2$ is the ratio of quark-gluon condensate to quark condensate 
introduced earlier in Eqs. (\ref{eq:theta}) and (\ref{eq:m0}). 
At this point our results are already slightly different from 
\cite{KKZ,KKY,Z}. Namely, we have 
included additional contributions related to the fact that 
the octet combination of color EDM operators has the quantum numbers of the
$\pi^0$ and $\eta$ fields which can therefore be produced from the vacuum.
$\pi^0$, for example, 
can be ``rescattered'' on the nucleon with an amplitude proportional to 
$(m_d+m_u)\langle N|\bar uu +\bar dd|N\rangle$. As a result, the diagram
shown in Fig. 2 is responsible for a contribution directly proportional to
$m^2$ which is effectively of the same order as the direct contribution
considered in
\cite{KKY,KK}. 
\begin{figure}[thb]
 \begin{center}
\epsfig{file=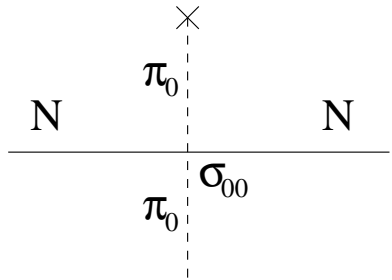,height=1.5in} 
%\vspace{-0.5in}
\caption{Additional contribution to the $\bar g_{\pi NN}$, proportional
to the nucleon sigma term}   \ \end{center}
\vspace{-0.3in}\end{figure}

Further calculation relies on QCD sum rules \cite{SVZ} and low-energy 
theorems in QCD. Matrix elements from  
$q g_s(G\mbox{$\sigma$})q$ operators were evaluated in \cite{KKZ,KKY,Z}:  
\begin{equation} 
\langle p|\bar q g_s(G\mbox{$\sigma$})q|p\rangle\simeq \frac{5}{3} m^2 
\langle p|\bar qq|p\rangle.  \label{eq:gluonium} 
\end{equation} 
The matrix 
elements over the proton can be obtained from baryon mass splittings and 
pion-nucleon scattering data. Here we take the following values for the 
matrix elements of $\bar qq$ over the nucleon \cite{Z}:  
\begin{equation} 
\langle p|\bar uu|p\rangle\simeq 4.8;\;\;\langle p|\bar dd |p\rangle\simeq 
4.1;\;\; \langle p|\bar ss|p\rangle\simeq 2.8  \label{eq:me} 
\end{equation} 
These values of $\bar qq$ matrix elements correspond to the choice 
$m_u=4.5$ MeV, $m_d=9.5$ MeV and $m_s=175$ MeV. The values of these 
matrix elements, together with the factorization formula (\ref{eq:gluonium}),
suggest that T-odd 
nucleon-nucleon forces are primarily 
sensitive to $\tilde d_u-\tilde d_d$ and insensitive to $\tilde 
d_u+\tilde d_d$, simply because the contribution to $\bar g_{\pi pp}$ 
proportional to $\tilde d_u+\tilde d_d$ in (\ref{eq:matrel}) is relatively
suppressed by
\begin{equation} 
\frac{\bar g_{\pi pp}(\tilde d_u+\tilde d_d)}
{\bar g_{\pi pp}(\tilde d_u-\tilde d_d) } \simeq 
\frac{2\langle p|\bar u u - \bar d d 
|p\rangle}{\langle p|\bar u u + \bar d d|p\rangle} \sim 0.2
\label{+/-}
\end{equation} 
In this sense, the contribution furnished by $\theta_{eff}$ is numerically 
insignificant because $\bar g_{\pi pp}$ generated by $\theta$ is also 
proportional to $\langle p|\bar u u - \bar d d |p\rangle$.
 
Thus, these simple considerations suggest that due to the  
numerical dominance of  the 
triplet combination of color EDM operators, the final answer for 
$\xi$ takes the following form:  
\begin{equation} 
\xi=G_F^{-1}\frac{3 g_{\pi pp}m_0^2}{f_\pi m_\pi^2} (\tilde d_d-\tilde 
d_u-0.012\tilde d_s),  \label{eq:xi} 
\end{equation} 
We can see that the contribution of the strange quark CEDM is numerically 
suppressed, mainly due to the additional smallness of $\eta NN$ 
CP-conserving interaction as compared to $g_{\pi NN}$. 
 
Combining equations (\ref{eq:Sexp}), (\ref{eq:Sth}) and (\ref{eq:xi}), we 
arrive at the following prediction for the EDM of the mercury atom:  
\begin{equation} 
d_{Hg}=-(\tilde d_d-\tilde d_u-0.012\tilde d_s)\times 3.2 \cdot 10^{-2} e,  
\label{eq:EDMhg} 
\end{equation} 
where the the numerical coefficient $3.2 \cdot 10^{-2}$ corresponds to the
choice of light quark masses given above.
Using the experimental limits (\ref{eq:Hglim}), we deduce the very strong 
constraint on the following combinations of the CEDMs of quarks:  
\begin{equation} 
| \tilde d_d-\tilde d_u-0.012\tilde d_s | < 3.0\cdot 10^{-26} cm.  
\label{lim} 
\end{equation} 

It is important to note that the quark 
EDM operators cannot induce a large value for $S$. They do not induce the
$\bar ni\gf n \bar pp$ interaction, and their contribution to the 
Schiff moment of the nucleus is associated only with electric dipole 
moment of the external valence nucleon \cite{KL}. Current limits on $d_{Hg}$
are only sensitive to quark EDMs larger than $ 10^{-24}e\cdot cm$ and thus 
these operators can be safely neglected.      
Similarly, the potential contribution from the three gluon operator $GG\tilde
G$ to $d_{Hg}$ are small. 
We rely here on the QCD sum rule estimates \cite{Volodia}, showing no 
significant 
contribution from this operator to the T-odd nucleon-nucleon forces and
thus to the EDM of mercury.  

Finally, we would like to comment on the accuracy of the predictions 
(\ref{eq:EDMhg}) and (\ref{lim}), distinguishing between the error
in the overall coefficients and the errors in the relative coefficients 
of $\tilde d_i$-proportional contributions. The uncertainties of the atomic 
calculations of $d_{Hg}(S)$ and nuclear calculations of $S(\xi)$ 
mostly affects the overall coefficients. Although the uncertainty in the overall
coefficient can be significant \cite{KL}, it 
is acceptable for our purpose as it influences only the width of the 
allowed region in $\theta_\mu-\theta_A$-plane. What is more important,
however, is that the prediction of the relative coefficients in front of
individual $\tilde d_i$ in eqs.(\ref{eq:EDMhg}) and (\ref{lim}) can be
done in a more reliable way and we estimate that the accuracy of keeping 
the triplet combination $\tilde d_d-\tilde d_u$ and neglecting 
$\tilde d_d+\tilde d_u$, eq. (\ref{+/-}), is at the level of 20\%. 
In effect, it makes the constraints imposed by $d_{Hg}$ much more useful than
those provided by $d_n$. Another advantage of the approach for calculating 
$d_{Hg}$ and $d_n$, developed in refs. \cite{KKY,KK,KL} and applied here, 
is that it reduces the error from the poor knowledge of the light quark masses.
Indeed, even in the case of the na\"{\i }ve formula for the EDM of the 
neutron, $d_n\simeq (4d_d-d_u)/3$, the individual quark EDM contributions 
are proportional to $m_{u,d}$ which are known to 50\%. In the present
approach, the answer for $\xi$ is ultimately proportional to a linear 
combination of $m_i\langle 0|\bar qq|0\rangle$, which can be rewritten 
as $f_{\pi}^2m_\pi^2$ times the function which depends only on the 
{\em ratio} of light quark masses, known to much better
accuracy than the masses themselves.

\section{The limits on the MSSM CP-violating phases} 
 
In previous work \cite{FO1}, limits on the neutron and electron dipole
moments were  used to constrain the two independent phases (of $\mu$ and
$A$) in the MSSM  assuming that all the terms in the Higgs potential and all
gaugino masses  are real and that all of the $A$-parameters are equal at the
GUT scale and  share a common phase. In absolute terms, the phases are not
overly  constrained, $\theta_\mu \stackrel{<}{{}_\sim} 0.3$, for $\theta_A
\simeq 
\pi/2$. The reason for the lax limits, are several cancellations in the 
various contributions to the EDMs. Furthermore, in some regions of parameter 
space, these cancellations occur simultaneously for the electron and neutron 
EDMs. 
 
As we argued above, there are several reasons to suspect that the limit due 
to the neutron EDM must be treated with caution. Instead, we have argued 
that the limit coming from the EDM of Hg is the result of a ``cleaner" 
calculation and carries fewer QCD uncertainties. In what follows, we will 
explore in detail the limits on the two phases using the constraint based on 
the EDM of $^{199}$Hg (\ref{lim}) derived above. We will compare these 
constraints on the phases to that obtained from the electron EDM. As we will 
see, the cancellations in the EDMs do not always occur at the same points in 
parameter space. To demonstrate the importance of the mercury  EDM limit, we 
first consider a SUSY model with a single mass scale. We then present 
general results which assume gaugino and sfermion mass universality at the 
GUT scale. 
 
Following \cite{FOS,FO}, we analyze the limits on $\theta_A$ and $\theta_\mu$
for  different values of supersymmetric parameters. To demonstrate the 
sensitivity of the mercury EDM to a common scale of the supersymmetric masses 
with arbitrary and uncorrelated phases, we choose $m_{\tilde f}\simeq 
M_{\lambda_i}\simeq |\mu|\simeq |A_k|$ at the electroweak scale and take $ 
\tan\beta=2$. In Figures 3a and 3b, we show the sensitivity to the EDM of the 
mercury atom for the cases of $\sin\theta_A=1,~\sin\theta_\mu=0$ and $ 
\sin\theta_A=0,~\sin\theta_\mu=1$. At this particular point of the 
supersymmetric parameter space all of the calculations are significantly 
simplified.  When all soft-breaking parameters are 
sufficiently heavy, close to the TeV scale, the chargino and gluino 
propagators can be simply expanded in $v_1/M$ or $v_2/M$ and only the zeroth
and  first order terms in the expansions need be kept. If needed, for lower
values of  gaugino masses, the results can be generalized to include all
effects of mixing in the gluino and chargino sectors.

\begin{figure}
\normalsize 
\begin{center}

%\vspace{-1cm} 

%\epsfig{file=r3.eps,height=100mm}
\mbox{\epsfxsize=100mm\epsffile{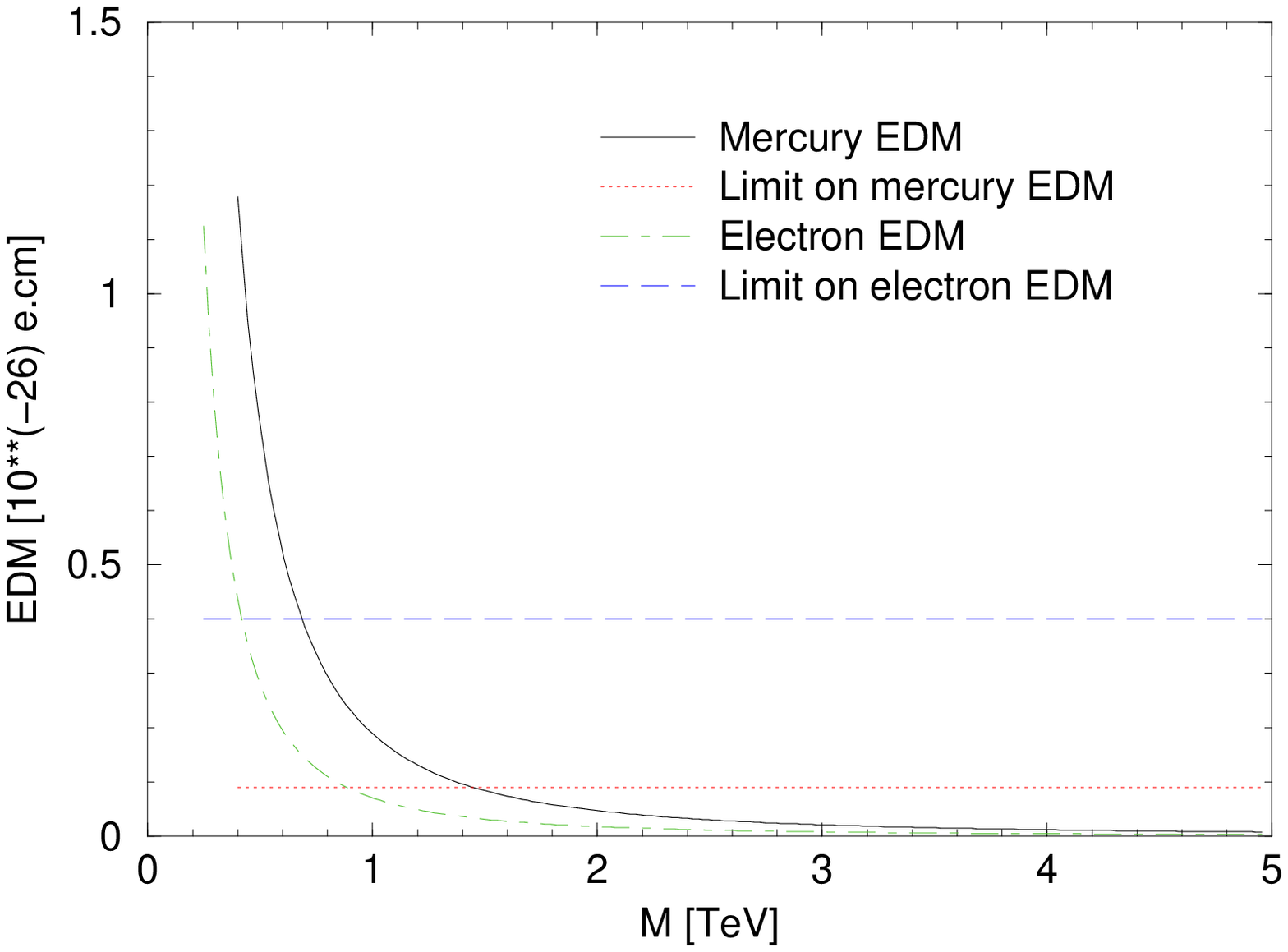}}

\vspace{1cm}

\mbox{\epsfxsize=100mm\epsffile{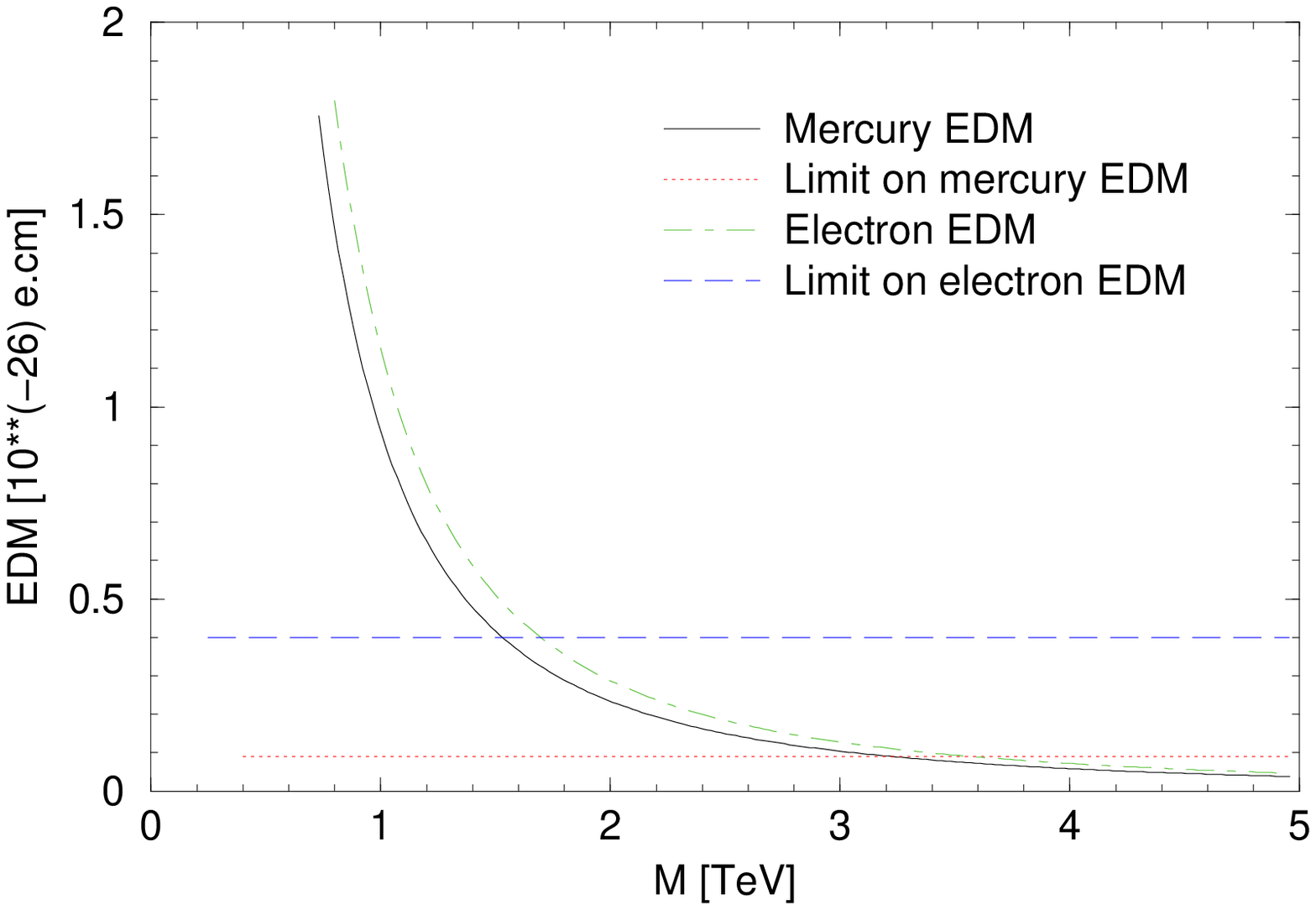}}

\vspace{1cm}
\end{center}

\caption{\label{fig:mssm}The sensitivity of the EDMs of mercury and electron 
   to the scale of the soft-breaking parameters with  a) maximal phase of
   $A$ ($\theta_A=\pi/2$, $\theta_\mu=0$), and b) maximal phase of
   $\mu$ ($\theta_A=0$, $\theta_\mu=\pi/2$).  We've taken
   $|\mu|=|A|=m_{\tilde Q}=m_{\tilde U}=m_{\tilde
     D}=M_{\lambda_i}\equiv M$.  The horizontal line is the current
   experimental limit.
 }

\end{figure}

The calculation of the chromoelectric dipole moments of quarks in MSSM was 
done in the series of papers \cite{Duff,KZ,IN}. When the CEDMs of quarks are 
induced by $\theta_A$, (as in Fig 3a), the result is dominated by gluino 
exchange, with very small corrections coming from $\lambda_1$-exchange:  
\begin{eqnarray} 
\tilde d_d=-\eta\frac{m_d|A|\sin\theta_A}{16\pi^2M^3} \left(\frac{5g_3^2}{18}-
\frac{g_1^2}{108}\right) \\ 
\tilde d_u=-\eta\frac{m_u|A|\sin\theta_A}{16\pi^2M^3} \left(\frac{5g_3^2}{18}+
\frac{g_1^2}{54}\right).  \nonumber 
\label{eq:phiA}
\end{eqnarray} 
Here $\eta$ denotes the renormalization group factor which reflects the 
QCD evolution of the color EDM from the weak scale to 1 GeV. When the color
EDM operator is defined as in eq. (\ref{eq:eff}), its anomalous dimension
is negative and small so that the overall renormalization of $\tilde d_i$ is
not important. The alternative definition of color EDM operator, 
frequently occurring
in literature, is $\frac{1}{2}\tilde{d}'\bar{q}t^{a}G_{\mu \nu }^{a}
\mbox{$\sigma$}_{\mu \nu }\mbox{$\gamma_{5}$} q$, where $g_s$ is included
in $\tilde{d}'$. Defined this way, this operator acquires a
renormalization factor roughly proportional to $g_s(1{\rm
GeV})/g_s(M_Z)\simeq 2$ which is smaller than the value 3.3 quoted
in \cite{Duff}. This is because in Refs. \cite{Duff,IN} a very large
coupling constant at low energies, $\alpha_s\simeq 2$, is used. There is,
however, an important numerical contribution to $\eta$ which  reflects the 
suppression of light quark masses at the high energy scale, 
$m_d(M_Z)/m_d({\rm 1GeV})$. We choose 
to use low energy values for $m_u$ and $m_d$, 4.5 and 9.5 MeV,
and include the quark mass RG factor into $\eta$. 
For the scale $M$ of order $M_Z$, $\eta$ is 
numerically close to 0.35 and is mainly due to the suppression of the quark 
masses at the high energy scale. This suppression factor was omitted
in Ref. \cite{Duff} where $m_u(M_Z)=8$ MeV is used.

Combining all numerical factors, we obtain the following value for the EDM 
of $^{199}$Hg:  
\begin{equation} 
d_{Hg}=e\cdot 1.5\cdot10^{-2}\frac{5\mbox{$\alpha$}_3}{72\pi} \frac{
(m_d-m_u-0.012m_s)|A|\sin\theta_A}{M^3}\simeq 2\cdot 10^{-27}
\left(\frac{{\rm 1 TeV}}{M}\right)^2~e\cdot cm, 
\end{equation} 
where we simply take $|A|=M$. We see from Fig. 3a that the mercury EDM
places a constraint on $M$, $M \ga 1.5$ TeV.

In the other case, with $\sin\theta_A=0,~\sin\theta_\mu=1$, we have to include
$\lambda_2$-higgsino and $\lambda_1$-higgsino exchanges as well, so that
the  result for the CEDMs, (shown in Fig. 3b), is as follows:  
\begin{eqnarray} 
\tilde d_d=\eta\frac{m_d|\mu|\tan\beta\sin\theta_\mu}{16\pi^2M^3} \left(\frac{ 
5g_3^2}{18}+\frac{g_2^2}{8}+\frac{g_1^2}{216}\right) \\ 
\tilde d_u=\eta\frac{m_u|\mu|\cot\beta\sin\theta_\mu}{16\pi^2M^3} \left(\frac{ 
5g_3^2}{18}+\frac{g_2^2}{8}+\frac{7g_1^2}{216}\right).  \nonumber 
\end{eqnarray} 
As a result, the contribution of the up quark relative to that of the down 
quark is suppressed by $m_u/(m_d\tan^2\beta)$. Numerically, the gluino 
exchange diagram dominates again with less than a 10\% contribution coming 
from $\lambda_2$-higgsino exchange. In this case, the limit is somewhat
stronger giving, $M \ga 3$ TeV.

As one can see, the EDM of mercury is sensitive to the scale of 
supersymmetric masses as high as 1.5-3 TeV. This can be compared with the 
sensitivity of the EDM of the electron, which we calculate at the same point 
of the supersymmetric parameter space, taking the slepton masses equal 
to the squark masses:  
\begin{eqnarray} 
d_e & = & \frac{m_e|A|\sin\theta_A}{16\pi^2M^3}\frac{g_1^2}{12} \\ 
d_e&  = & \frac{m_e|\mu|\tan\beta\sin\theta_\mu}{16\pi^2M^3}
\left(\frac{5g_2^2}{24}+
\frac{g_1^2}{24}\right).  
\nonumber
\label{eq:EDMe} 
\end{eqnarray} 
The limits based on the electron EDM, for the two cases
considered, are weaker as can be seen from Fig. 3a and 3b
where the limit on $M$ is 0.4 and 1.7 TeV.

There is also the possibility 
of destructive interference between two contributions induced by the
CP-violating phases. Again, we choose the supersymmetric parameters to be 
equal and fix them to be in the range from $250$ -- $750$ GeV. 
Fig. 4a-4c
show the combined exclusion plots. The two bands correspond
to the  parts of the parameter space where the mercury or electron (Tl)
constraints are lifted by the cancellation of different supersymmetric
contributions.  The allowed area lies on the intersection of these two
bands. We observe that the band corresponding to the mercury EDM
constraint has a different  slope than that of the electron EDM, mainly
because $d_{Hg}$ is by far more sensitive to $\theta_A$. We observe that
{\em both} phases are sufficiently  constrained for the low values of
$M$.

\begin{figure}[hbtp]

 \vspace{0cm}

\begin{center}
\hspace*{-1.3cm}\normalsize \mbox{\epsfxsize=70mm\epsffile{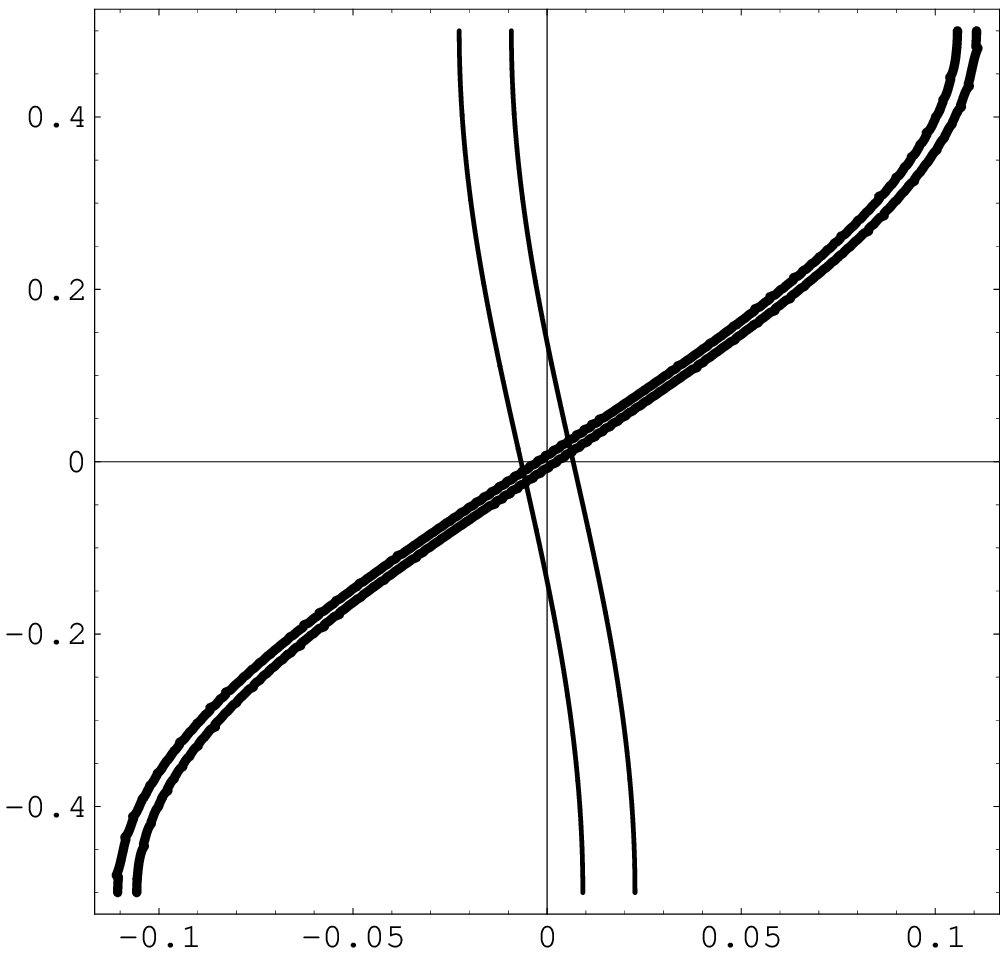}}
\hspace*{2cm}\normalsize \mbox{\epsfxsize=70mm\epsffile{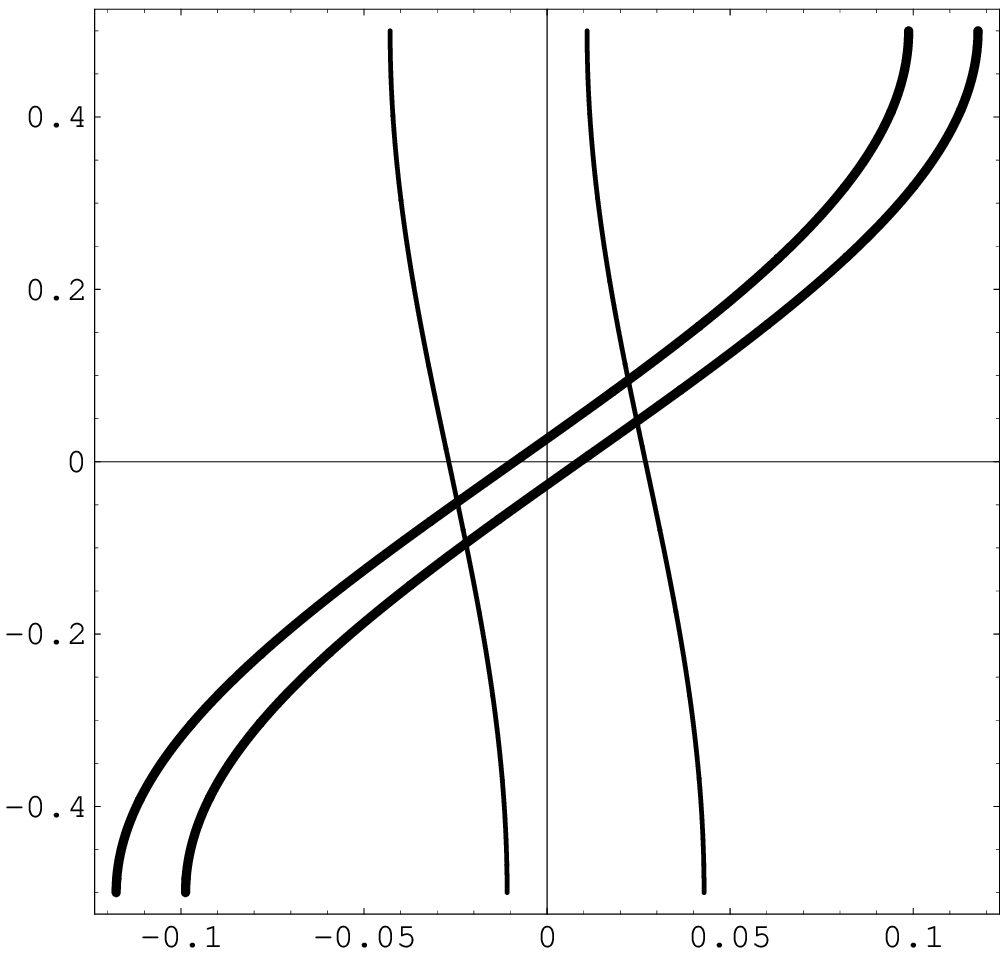}}  

\vspace{-4.3cm} 

\large 
\hspace{-8.5cm} $\frac{\theta_A}{\pi}$ \hspace{-10cm} $\frac{\theta_A}{\pi}$

\vspace{3.5cm} 

\normalsize

\vspace*{-3cm}
\hspace*{.1cm}$d_{Hg}$ \hspace*{8.1cm} $d_{Hg}$\hspace*{5.2cm}
\vspace*{2.4cm}

\vspace*{-2cm}
\hspace*{2.2cm}$d_{e}$ \hspace*{9cm} $d_{e}$
\vspace*{1.4cm}

\hspace{-1cm} $\theta_\mu/\pi$ 
\hspace{-10.3cm} $\theta_\mu/\pi$

%\vspace{2cm}

\end{center}

 \hspace*{1.5cm}\normalsize Fig. 4a: $M$ =250 GeV 
\hspace*{4.6cm}  Fig. 4b: $M$ =500 GeV 

\vspace{1cm}

\end{figure}
\newpage
\begin{figure}
\begin{center}
\normalsize \mbox{\epsfxsize=70mm\epsffile{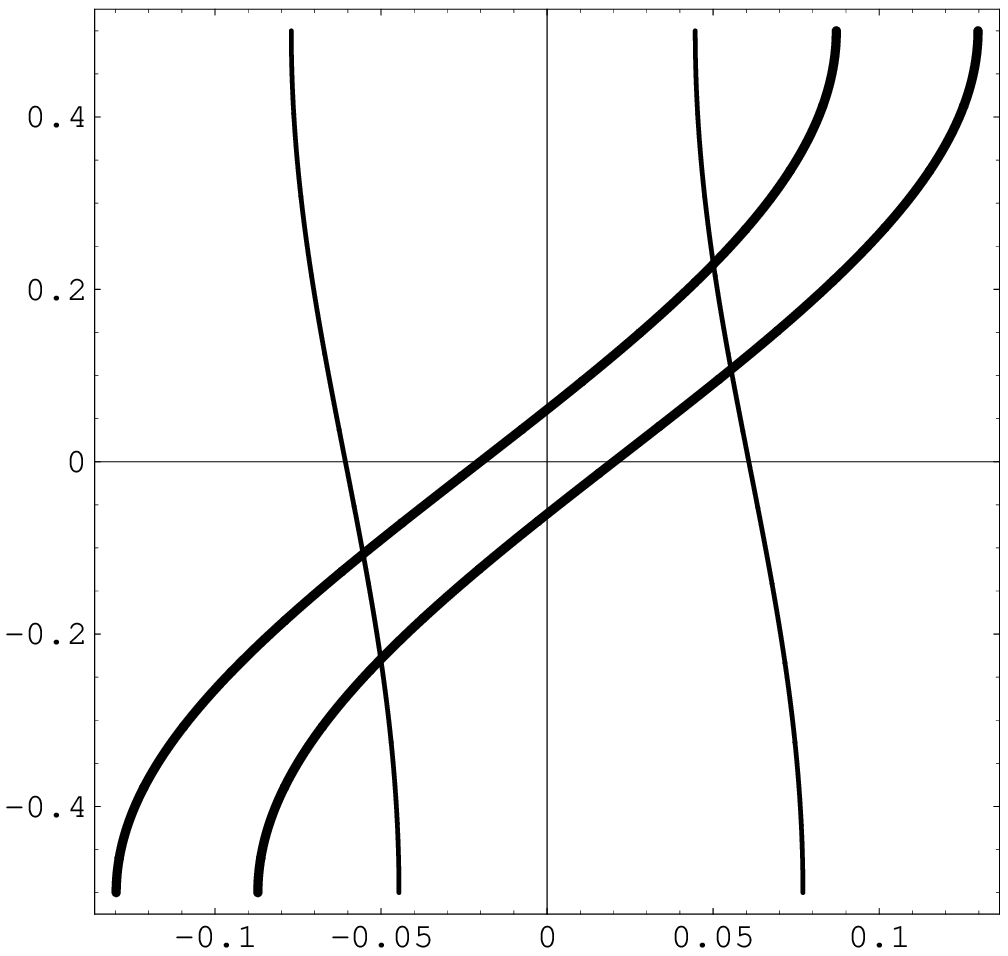}}  

\vspace{-4.4cm} 
\large 
\hspace{-7.5cm} $\frac{\theta_A}{\pi}$

\normalsize
\vspace*{+1cm}
\hspace*{.1cm}$d_{Hg}$ \hspace*{4.6cm}
%\vspace*{0.4cm}

%\vspace*{-2cm}
\hspace*{2.2cm}$d_{e}$ %\hspace*{9cm}
%\vspace*{1.4cm}

\vspace*{-2cm}

\vspace{3.7cm}\hspace{0.5cm} $\theta_\mu/\pi$ 

\vspace{1cm}

\normalsize Fig. 4c: $M$ =750 GeV 
\end{center}
\caption{
Combined, $d_{Hg}$ and $d_e$, constraints on the supersymmetric phases $ 
\theta_A/\pi$ and $\theta_\mu/\pi$ for different values of $M$. Allowed
area is on the intersection of two bands.}

\end{figure}

\section{EDMs in mSUGRA and Cosmological Constraints}

\begin{figure}[thb]
\begin{center}
\epsfig{file=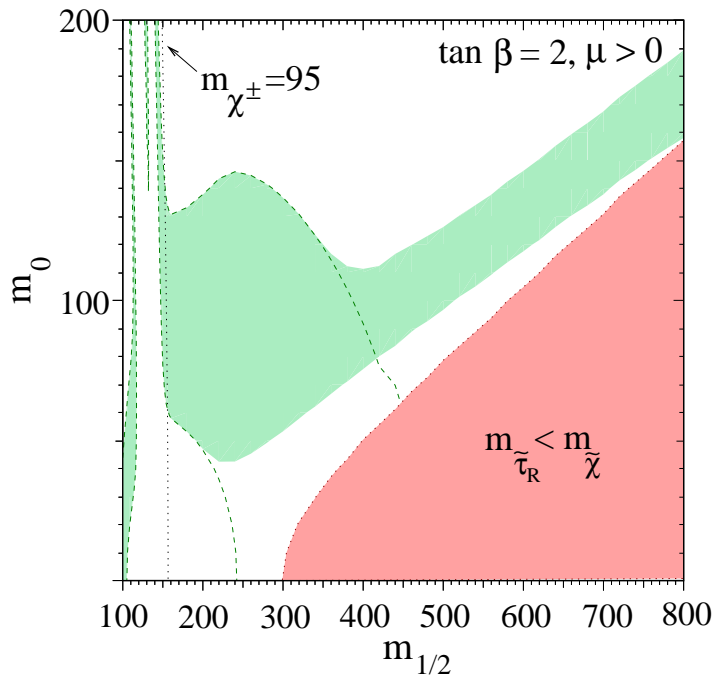,height=3.5in} 
%\vspace{-0.5in}
\caption{\label{fig:rd}The light-shaded area is the cosmologically preferred 
  region with \mbox{$0.1\leq\ohsq\leq 0.3$}.  The dashed line shows
  the location of the cosmologically preferred region if one ignores
  the light sleptons.  In the dark shaded region the LSP is the
  ${\tilde \tau}_R$, leading to an unacceptable abundance of charged
  dark matter.  Also shown as a dotted line is the contour $m_{\chi^{\pm}}=95\gev$.}  
\end{center}
\vspace{-0.1in}\end{figure}

We now consider the constraints on $\cp$ violating phases in
mSUGRA-like models, i.e. models with unified gaugino and sfermion
masses.  We recall that to one loop, the phase of $\mu$ does not
evolve with scale, but the phases of $A_u, A_d$ and $A_e$ must be run
separately from the unification scale to low energies.  We follow the
analysis of \cite{FO1,FO}, but with two changes.  First, we replace
constraints from the neutron electric dipole moment with limits from
the EDM of Hg, discussed above.  Second, we include recent results on
the effect of coannihilations of neutralinos with staus on the
neutralino relic density \cite{EFO}.  The latter has the effect of
weakening the cosmological upper bound on the gaugino masses.  This is
demonstrated in Fig.~\ref{fig:rd}, where the light shading indicates
the region of the $\{m_0,\m12\}$ plane which yields a neutralino relic
abundance in the cosmologically preferred range $0.1\le\ohsq\le0.3$.
The upper limit of the light shaded region crosses below the
line $\mchi=m_{\st}$ at $\m12\sim1400\gev$; for greater $\m12$, either
the relic density violates the upper bound $\ohsq\le 0.3$ (which follows
from a lower limit of $12 \gyr$ on the age of the universe) or the
lightest supersymmetric particle is a stau, leading to an unacceptable 
abundance of charged dark matter.
Here we've taken $\tb=2$, but the light shaded region is quite
insensitive to $\tb$ for the values of $\tb$ we consider, as well as 
insensitive to the
phase of $\mu$\cite{FO}.  For
comparison, the dashed lines demarcate the inferred cosmologically
preferred region if one ignores the effects of neutralino-slepton
coannihilation.  Whereas in \cite{FO1}, the constraint $\ohsq\le0.3$
yielded an upper bound of $450\gev$ on $\m12$, we now have to consider
larger values of $\m12$.  However, we will see that this does not
effect the upper bound on $\thm$.

In contrast to the results of the previous section, we find that in
mSUGRA-like models, constraints from the electron EDM are typically
more restrictive than those from the EDM of Hg.  This difference
arises because in models with gaugino masses unified at the GUT scale,
the gluino tends to be considerably heavier than the neutralino and
charginos, and this suppresses the contribution to $d_{Hg}$ from the
quark chromoelectric dipole moments due to gluino exchange.  We recall
that cancellations between the chargino and neutralino exchange
contributions to the electron EDM allow for large values of
$\thm$ \cite{FO,IN,FO1}.   A similar effect also applies in the case of the
Hg EDM, where cancellations can occur between the gluino exchange and
neutralino and chargino exchange contributions to the quark
chromoelectric dipole moments.  The power of combining the electron and 
Hg limits lies in the fact that for fixed $\thm$ and $\tha$, the 
cancellations in
the electron and Hg dipole moments occur for different, and often
non-overlapping, ranges in $\m12$.  Thus the combined limits are
stronger than either limit alone.

\begin{figure}
\vspace*{-2.3in}
\begin{minipage}{6.0cm}
\hspace*{-1in}
\epsfig{file=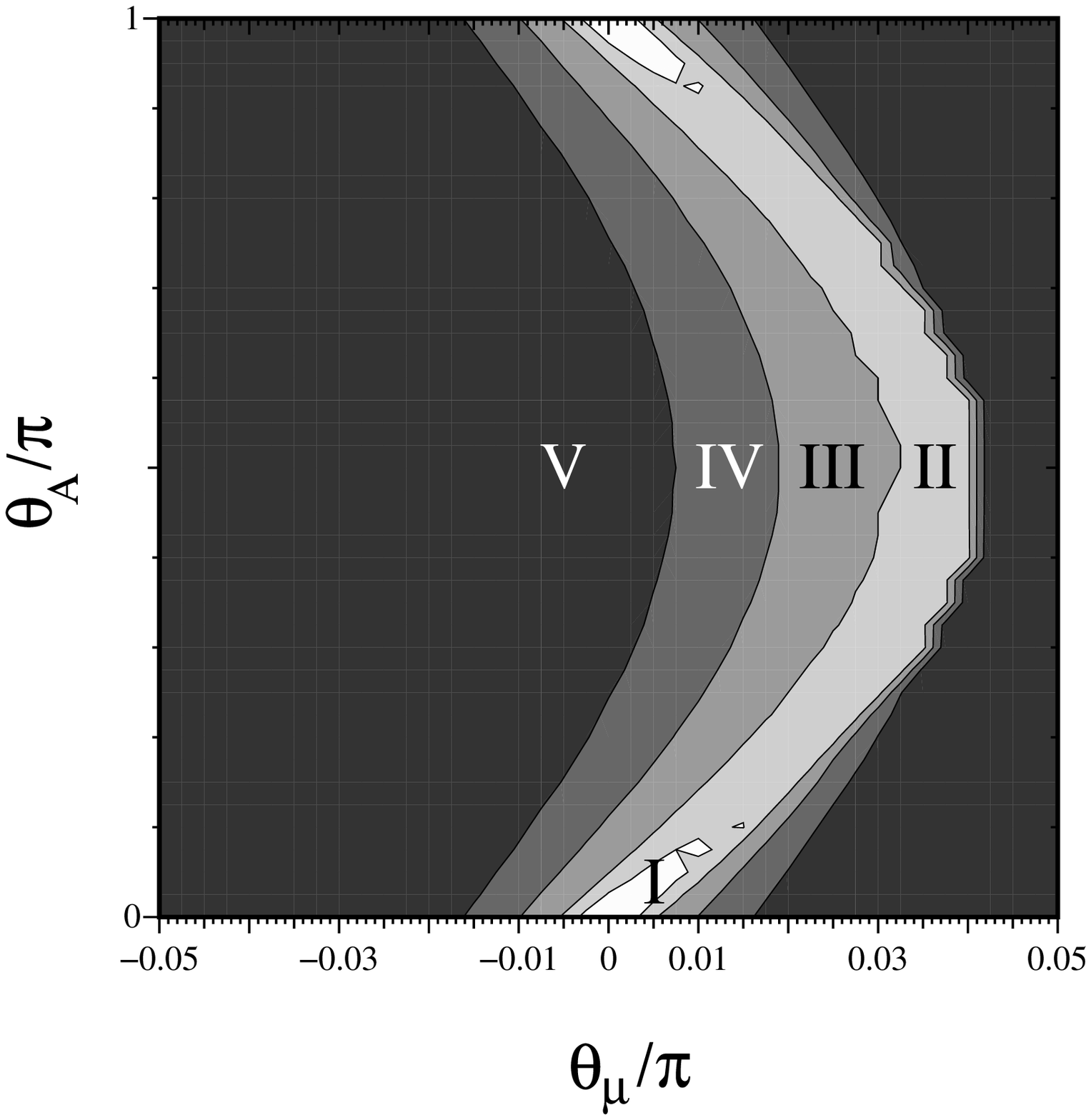,height=6in} 
\end{minipage}
\hspace*{0.3in}
\begin{minipage}{6.0cm}
\epsfig{file=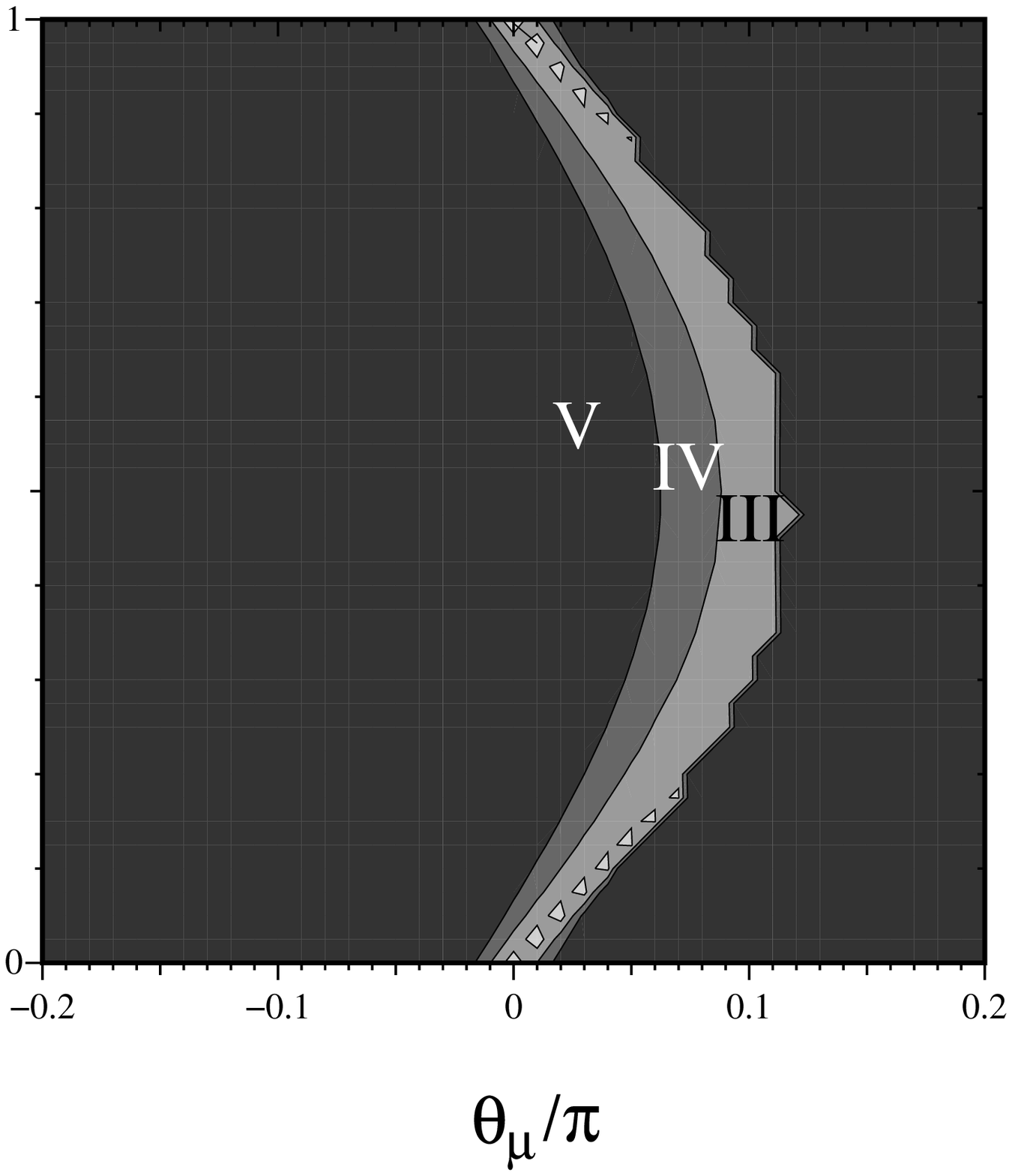,height=6in} 
\end{minipage}\hfill
\vspace{-2.3in}
\begin{minipage}{6.0cm}
%\hspace*{-1in}
\hspace*{-1in}
\epsfig{file=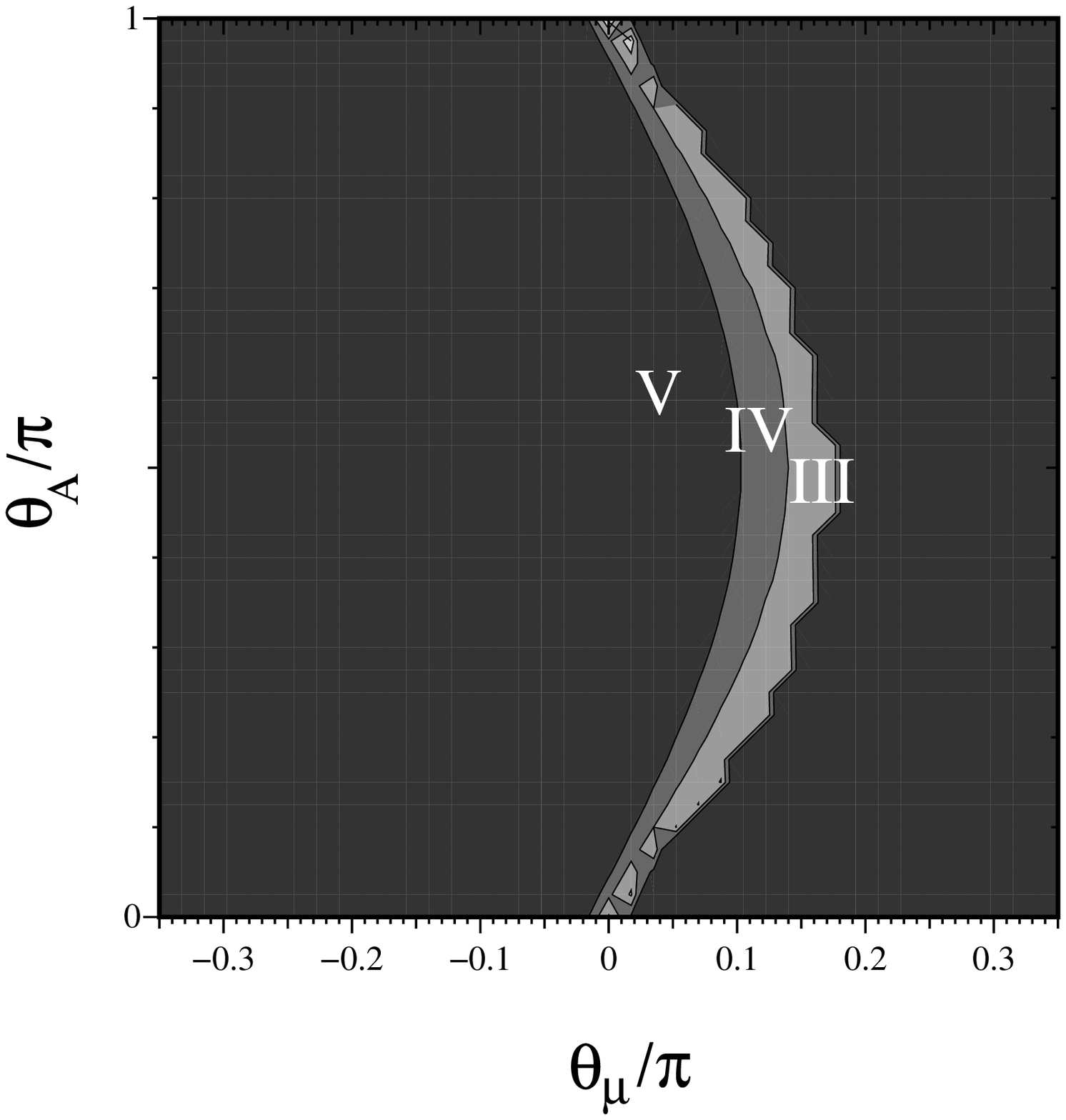,height=6in} 
\end{minipage}
\hspace*{1.0in}
\begin{minipage}{6.0cm}
\vspace*{0.65in}
\epsfig{file=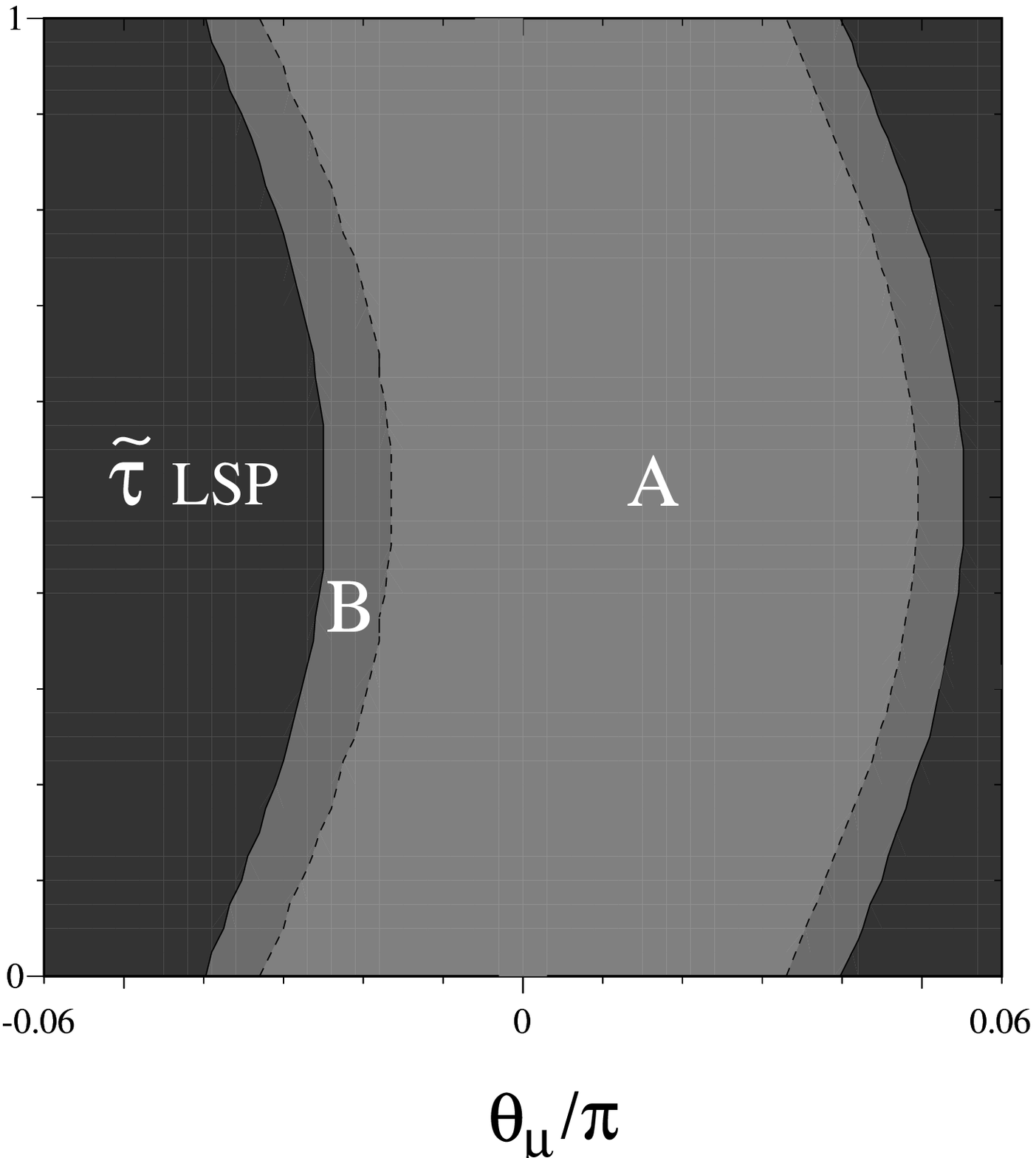,height=3.5in} 
\end{minipage}\hfill
\vspace{-0.7in}
\caption{\label{fig:bedm}Contours of $\m12^{\rm min}$, the minimum $\m12$ 
  required to bring both the electron and Hg EDMs below their
  respective experimental bounds, for $\tan\beta=2, m_0=130\gev$, and
  a)$A_0=300\gev$, b)$A_0=1000\gev$ and c)$A_0=1500\gev$.  The central
  light zone labeled ``I'' has $\m12^{\rm min}<200\gev$, while the
  zones labeled ``II'', ``III'', and ``IV'' correspond to
  \hbox{$200\gev<\m12^{\rm min}<300\gev$}, $300\gev<\m12^{\rm
    min}<450\gev$, $450\gev<\m12^{\rm min}<600\gev$ and $\m12^{\rm
    min}>600\gev$, respectively.  Zone V is therefore cosmologically
  excluded. Panel d) shows the allowed region in the ``trunk'' for
  $A_0=300\gev, m_0=200\gev$.  The light shaded regions ``A'' and
  ``B'' are permitted, whereas the dark shaded region is
  cosmologically excluded (see the text).}
\end{figure}

Following \cite{FO,FO1}, we compute the electron and Hg EDMs in mSUGRA
as a function of $\thm,\tha$ and $\m12$ for fixed $A_0,m_0$ and $\tb$.
In Fig.~\ref{fig:bedm}a-c we display the
minimum value of $\m12$ required to bring both the electron and Hg
EDMs below their experimental limits, for $\tb=2$ and $m_0=130\gev$.  We
exclude points which violate the current LEP2 chargino and slepton mass
bounds \cite{expt}.  The EDMs are computed on a 40x40 grid in
$\{\thm,\tha\}$, and features smaller than the grid size are not
significant.  Although the dependence of the EDMs on $\m12$ is not
monotonic, there is still a minimum value of $\m12$, due to cancellations,
which is permitted.  In the zone labeled ``I'', $\m12^{\rm min}<200\gev$,
while the zones labeled ``II'', ``III'', ``IV'' and ``V'' correspond to
$200\gev<\m12^{\rm min}<300\gev$, $300\gev<\m12^{\rm min}<450\gev$,
$450\gev<\m12^{\rm min}<600\gev$ and $\m12^{\rm min}>600\gev$,
respectively.  Comparing with Fig.~\ref{fig:rd}, we see that values of
$\m12$ larger than about $600\gev$ are cosmologically excluded for this
value $m_0$.  Therefore, region V corresponds to an excluded region in the 
phase plane.    Of course, for this value of $\tb$, the current Higgs mass bound
requires enormous sfermion masses $\gg 1\tev$, which are cosmologically 
prohibited.  We've chosen to plot our results for $\tb=2$ in order to compare to 
our previous results \cite{FO,FO1}.  Qualitatively similar conclusions
apply for larger $\tb$, which we summarize at the end of this section.

Figure 2 of Ref.~\cite{FO1} displays the corresponding contours to our
Fig.~\ref{fig:bedm}a-c, but imposing only the constraint from the
electron EDM\footnote{In \cite{FO1} we take $m_0=100\gev$, rather than
  $130\gev$; however, taking  $m_0=130\gev$ makes only a small change 
in the displayed
  contours and a slight reduction in the upper bound on $\thm$. }.
Note that in \cite{FO1} we do not include a contour corresponding to
$450\gev<\m12^{\rm min}<600\gev$, as this region would be
cosmologically excluded in the absence of coannihilations of
neutralinos with sleptons, whose effects were not included in
\cite{FO1}. The effect of including the Hg EDM bounds is
particularly significant at large $A_0$, where the cancellations are
enhanced and the bounds on $\thm$ are weakest.  Here the widths of the
allowed region in $\m12$ at fixed $\tha$ and $\thm$ are narrowest,
leaving less opportunity for overlap between the ranges allowed by the
electron and Hg EDMs, respectively.  Indeed, for $A_0=1.5\tev$, the
upper bound on $\thm$ is reduced from $\sim0.3\pi$, in the case of the
electron EDM alone, to $\sim0.18\pi$, combining the two constraints,
and, further, the width of the region in $\thm$ is considerably
narrowed.  The reduction in the bound on $\thm$ is minimal for small
$A_0$, where the bounds on $\thm$ are strongest.  However, notice that
the $\m12^{\rm min}$ at the largest allowed values of $\thm$ is
shifted from less than $200\gev$ in the case of the electron EDM alone
to between 200 and 300$\gev$ for the combined bound, in the case
$A_0=300\gev$.  For $A_0=1\tev$ and $1.5\tev$, $\m12^{\rm min}$ lies
above $300\gev$ at the largest $\thm$.

We note that the larger values of $\m12$ which neutralino-slepton
coannihilation permit do not increase the maximum $\thm$, for this
value of $m_0$.  This is
because, as we see from Fig.~\ref{fig:bedm}, the region of mutual
cancellations happens to lie at lower $\m12$, between 300 and
400$\gev$.  The widths of the allowed regions in $\m12$ are typically
between 50 and 80$\gev$ for the lightest shaded zones in
Fig.~\ref{fig:bedm}b and \ref{fig:bedm}c and greater than 80$\gev$
almost everywhere in Fig.~\ref{fig:bedm}a.  Larger $\m12$ does,
however, widen the allowed swath in $\thm$, by the region labeled
``IV'' in Fig.~\ref{fig:bedm}.  It helps in particular at small
$\thm$, where the electron EDM can be beaten down sufficiently by
taking heavy gaugino masses, without resorting to cancellations
between different contributions.  At large $\m12$, the Hg EDM
typically provides little constraint on $\thm$ due to the heaviness of
the gluinos.  As $m_0$ is increased, the regions of cancellation
shift, and the maximum value of $\thm$ slowly decreases.

\begin{figure}[thb]
 \begin{center}
\vspace*{-0.2in}
\epsfig{file=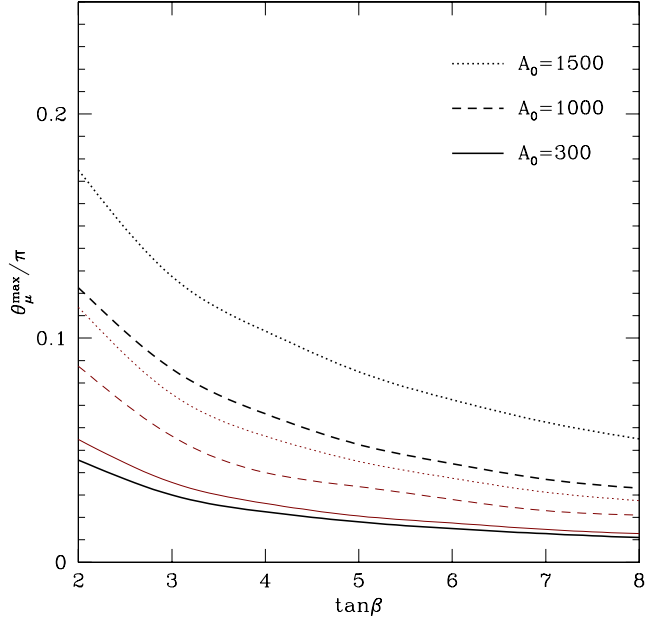,height=3.5in} 
\caption{\label{fig:thvtb}The maximum values of $\theta_\mu$ allowed by 
cosmology
  and both the electron and Hg EDMs, as a function of $\tan\beta$,
  for $m_0=100\gev$ (thick lines) and $m_0=200\gev$ (thin lines)
  and for $A_0 =300, 1000$ and $1500\gev$.}  
\end{center}
\end{figure}

These effects are enhanced at larger $m_0$ and $\m12$, in the
cosmologically allowed ``trunk'' which lies on top of the
$\tilde\tau_R$ LSP region (see Fig.~\ref{fig:rd}).  The allowed region
narrows as $m_0$ increases, and the $\ohsq=0.3$ contour crosses the
line $m_{\tilde\tau_R}=\mchi$ and gives an upper bound on $\m12$ and
$m_0$ at $\m12\sim1400\gev, m_0\sim300\gev$.  The trunk region yields
much larger sparticle masses than are cosmologically permitted in the
absence of coannihilations, and this can suppress the contributions to
the electric dipole moments sufficiently so that significant
cancellations between the various contributions are not necessary.
For low $A_0$, where the bounds on $\thm$ are tightest, the bounds on
$\thm$ are somewhat relaxed in the trunk area.  In
Fig.~\ref{fig:bedm}d, we display the allowed region in the
$\{\thm,\tha\}$ plane for $A_0=300\gev, m_0=200\gev$.  For this value
of $m_0$, $\m12$ is cosmologically restricted to lie between $850\gev$
and $950\gev$.  In the light region labeled ``A'', the EDMs are below
the experimental limits for all $850\gev\le\m12\le950\gev$, while in
the regions labeled ``B'', only part of this range of $\m12$
satisfies the EDM constraints.  The dark regions at large $|\thm|$
require $\m12>950\gev$ to satisfy the EDM bounds, and so these regions
are cosmologically excluded, as they yield a stau LSP.  The upper bound
on $\thm/\pi$ is relaxed to $\sim 0.055$.  Taking $\m12$ and $m_0$ at
their maximal values allows $\thm/\pi$ up to about 0.1.  

For large $A_0$, where $\thm$ can take its maximal values, the bound
on $\thm$ does not weaken in the trunk region.  As above, this is due
to the fact that at larger $\thm$, cancellations are still required to
bring the EDMs below their experimental limits, and the regions
of cancellations occur at lower $\m12$.  Even taking $\m12$ and $m_0$
at their largest cosmologically permitted values does not allow for
$\thm$ larger  than the bounds in Fig.~\ref{fig:bedm}b,c.  Further, since
the regions of low $\m12$ are cosmologically forbidden at large $m_0$,
the bounds on $\thm$ at large $A_0$ actually decrease for large $m_0$.
Thus the presence of the coannihilation trunk region does not increase
the overall combined cosmology/EDM bound on the phase $\thm$.

Lastly, we plot in Fig.~\ref{fig:thvtb} the maximum value of $\thm$
allowed by the electron and Hg electric dipole moments and the upper
limit on $\ohsq$, as a function of $\tb$.  The thick lines are for $m_0=100\gev$,
while the thin lines are for $m_0=200\gev$ and show the effect on the bounds
described above as one moves into the trunk region.

\section{Conclusions} 
 
We have shown that the calculation of the EDM of the neutron as the function 
of different MSSM phases is problematic due to large uncertainties 
related to the contributions of the color EDMs. 
 This is in contrast to the electric dipole moment 
of the mercury atom, induced by the T-odd nucleon-nucleon interaction. In 
the chiral limit the coefficient $\xi$, characterizing the strength of 
T-odd forces  has
a power-like singularity $\sim m_\pi^{-2}$, whereas $d_N \sim \log
m_\pi^{2}$ in the same chiral approach. 
It is apparent that the $\pi^0$ and $
\eta$ exchange diagrams dominate both parametrically and numerically and 
therefore yield a very good approximation to the magnitude of the T-odd 
interaction. The final result is proportional to $(\tilde d_d-\tilde 
d_u-0.012\tilde d_s)\times 3.2 \cdot 10^{-2} e$ and can be further
developed  in terms of CP-violating phases of the MSSM. 
 
There are two serious problems with the calculation of the T-odd nuclear 
forces due to the effective interaction (\ref{eq:eff}) with the coefficients 
provided by the MSSM. First is the status of the factorization in Eq. (\ref 
{eq:gluonium}), related with the low-energy theorem in $0^+$ channel. 
Following Refs. \cite{KKZ,KKY,KK}, we have have taken $\langle p|\bar q g_s(G
\mbox{$\sigma$})q|p\rangle\simeq 1.3 {\rm GeV}^2 \langle p|\bar qq|p\rangle$. 
We note that a designated sum rule calculation of this quantity and/or its 
simulation on lattice is highly desirable for it is the main source of 
uncertainties in the calculation of T-odd nuclear forces. The second 
potentially troublesome point is the effective negative sign between the
$\tilde  d_d$ and $\tilde d_s$ contributions. Although the numerical
suppression in  front of $\tilde d_s$ is quite strong and $d_d$ dominates,
destructive interference is still possible in both cases,
$\theta_A\neq 0$ and $ 
\theta_\mu \neq 0$. 
 
In this paper, we have considered first 
a very specific 
part of the supersymmetric parameter space, when all squark, slepton and 
gaugino masses were chosen to be equal. The theoretical prediction for $
d_{Hg}$ exhibits remarkable sensitivity to the scale of soft-breaking mass 
parameters as high as 1.5-3 TeV. When the scale is fixed below 1 TeV, $d_{Hg}$ 
limits {\em both} phases. The constraints on the CP-violating supersymmetric 
phases, obtained in this way are the strongest constraints so far. 

We have also considered the combined constraints from the Hg and electron
EDMs in the mSUGRA, when all supersymmetry breaking gaugino masses, 
soft scalar masses, and soft trilinear terms are separately unified at the
GUT scale. In this case, the sensitivity to the Hg EDM is weakened due to
the relative size of the gluino mass.  Nevertheless, the results are as
strong or stronger (particularly when $|A|$ is large and the limits are
weakest) than the combined results from $d_e$ and $d_N$.  The improvement in
the limit is due to the fact that cancellations among the contributions to
the EDMs occur at slightly different regions of the SUSY parameter space.

\section{Acknowledgments} 
 
M.P. would like to thank I.B. Khriplovich, A. Ritz, A.I. Vainshtein and A.R. 
Zhitnitsky for numerous important discussions. The work of M.P. and K.O. was
supported in part by DOE grant DE--FG02--94ER--40823.  The work of T.F. was
supported in part by DOE grant DE--FG02--95ER--40896 and in part by the
University of Wisconsin Research Committee with funds granted by the
Wisconsin Alumni Research Foundation.

\end{document}